# Reply to comment on "Attosecond electron microscopy and diffraction"

Dandan Hui, Husain Alqattan, Mohamed Sennary, Nikolay V. Golubev and Mohammed Th. Hassan.

Department of Physics, University of Arizona, Tucson, AZ 85721, USA.

## Introduction

### Attosecond temporal resolution in electron microscopy

Over the past few decades, following the first demonstration of ultrafast electron microscopy, numerous research groups have focused on achieving attosecond temporal resolution in electron microscopy with the goal of imaging electron and atomic motion. Recently, several studies have claimed to achieve attosecond temporal resolution in imaging, including reports by Nabben et al. (Nature 2023) (*1*), Gaida et al. (Nature Photonics 2024) (*2*), and Morimoto et al. (Nature Physics 2018) (*3*). These claims are based on the generation of attosecond electron pulse trains. However, in typical time-resolved measurements used to capture dynamic processes in real time, the temporal resolution is determined by the envelope of the pulse train. For example, in the work of Nabben et al. (Nature 2023) (*1*), the temporal resolution was effectively infinite because they used a continuous-wave (CW) laser pulse. In Gaida et al. (Nature Photonics 2024) (*2*), the resolution was limited by the 600-fs duration of the exciting pulse envelope as mentioned in (*2*), although it was mentioned 1.8 ps in the arXiv version! So, we are not sure which information is correct! Furthermore, the author is not providing raw data (only processed data) to confirm this information.

The reliance of using attosecond electron pulse trains fails to account for the distinct temporal resolution advantages enabled by our attosecond optical gating, which are absent in the case of using a continuous-wave or long laser pulse. These oversights highlight the limitations of this methodology (*1-3*), in studying ultrafast phenomena of matter.

***It is crucial to clarify this distinction to avoid confusion, misinterpretation, and potential miscitations within the community regarding attosecond temporal resolution in electron microscopy and the attosecond imaging of matter dynamics***.

To illustrate, in Nabben et al. (Nature 2023) (*1*), titled "*Attosecond electron microscopy of sub-cycle optical dynamics*" (along with other works from the same group based on similar principles), the authors use a CW laser pulse to shape free electrons from a transmission electron microscope (TEM) in an oscillatory form. They then use the same CW laser to excite scattered light from nanostructures. By adjusting the delay between the two beams with a highly precise piezoelectric stage (capable of nanometer-scale movement, corresponding to a few hundred attoseconds in light travel time), the authors *claim to achieve attosecond temporal resolution in electron microscopy and highlighted in the title which is misleading and overstatement*. The experiment is essentially a light interference setup, akin to observing interference fringes by delaying one CW laser pulse relative to another. *The resolution in this case is determined by the delay stage and the frequency oscillations of the laser pulses*, <u>not by the time response of the microscope which remains infinity</u>. Since this is not a true time-resolved measurement employing the "time-freezing" concept, it cannot be considered an attosecond experiment. In other words, the results in Nabben et al. (Nature 2023) (*1*) could be obtained using a TEM in energy-filtering mode, where two CW lasers illuminate a nanostructure, and their delay is controlled with attosecond precision to image the space-time propagation of light on this nanostructure—<u>*without the need to generate attosecond*</u>

*pulse trains*. This raises the question: **is this work a realistic demonstration of attosecond temporal resolution in electron microscopy?** In contrast, Hui et al. ([Science Advances 2024](#)) (*4*) presents the ***first realistic demonstration of attosecond imaging resolution in electron microscopy***, enabling the diffraction imaging of electron motion dynamics in graphene.

In a commentary by Peter Baum and Claus Ropers, the authors conjecture that the graphene dynamics observed in our time-resolved diffraction experiment (Fig. 5, Hui et al. 2024) (*4*) is an optical interference "artifact" or light modulation of electrons effects, similar to what was reported previously in Nabben et al. (Nature 2023) (*1*), Gaida et al. (Nature Photonics 2024)(*2*), and Morimoto et al. (Nature Physics 2018) (*3*), in addition to raising other technical concerns.

In this reply, we are pleased to address these allegations and provide clarifications to resolve the raised technical questions focusing on the published data on Hui et al. (*4*), since M. Yuan, et. al., arXiv:2411.02731 (2024) is unpublished data yet.

## 1. Addressing the Claim of "Lack of Electron Gating"

The first presumption by the Baum and Ropers, **"Lack of electron gating"**, is based on authors' statement *"Modulation without filtering is not gating"*. However, this reasoning is fundamentally flawed. The inability to detect directly a phenomenon due to a lack of the desired tools does not imply that the phenomenon does not exist.

To draw a parallel from attosecond science: XUV attosecond pulses, which were generated through high harmonic generation (HHG), were first created in the 1990s, but were not measurable until the streaking technique was developed in 2001—a breakthrough that ultimately earned the Nobel Prize in Physics in 2023. Before this development, researchers could not trace electron motion in real time, but this didn't imply that electrons weren't moving. The same logic applies here: the fact that the electrons in our experiment are not "filtered" in the traditional sense does not mean they are not gated or can't carry the signature of dynamics. The gated electrons in our setup, still capture the relevant dynamics, and the ungated electrons can be treated as background noise since they are blind to these dynamics and can be subtracted to isolate the dynamics of interest. This is the method we employed in Hui et al. (*4*).

## 2. Clarifying the "Incorrect Modulation Principle"

Baum and Ropers also raise concerns about the modulation principle we used, specifically comparing it to the polarization gating principle employed in HHG for generating isolated XUV pulses (*5*). This comparison is not appropriate. In our experiment setup, the evanescent electromagnetic fields generated by the leading circularly polarized part of the optical gating pulse (OGP) are separated in time by π from the fields produced by the trailing part of the pulse. These two fields are in opposite circular polarization directions (clockwise and anticlockwise), and when they are integrated on the mesh, they effectively cancel each other out, leaving only the plasmon field generated by the linear portion of the OGP. This is quite different from the HHG process, and Baum and Ropers's suggestion that our experiment follows this principle is, therefore, incorrect.

## 3. Addressing the Alleged "Optical Interference Artifacts"

The main contention raised by Baum and Ropers is the possibility that the observed dynamics are simply optical interference artifacts. However, we have experimentally demonstrated that this is not the case, and we provide evidence to refute this claim.

When we performed the diffraction experiment, we realized a scattered background. We subtracted this background before we started to record the data.

Then, upon reviewing the measured diffraction data at the time of the experiment (presented in Fig. 5, Hui et al. (*4*)), we noticed unexpected modulations in the background area surrounding the beam blocker. This prompted us to ask two critical questions:

**(i) What is the source of these background modulations?**

**(ii) How do these modulations affect the measured diffraction dynamics signals from the first, second, and third-order peaks in our measured data?**

To answer the first question and to investigate the origin of the background modulation further, we conducted a follow-up measurement (after recording our measurements at the same exact condition) where we blocked the UV beam to prevent electron generation inside the microscope, effectively eliminating any diffraction pattern. In this scan, we observed background modulations around the beam blocker area, which we identified as interference effects from the scattered lasers interacting on the beam blocker (the blocker is typically used in any diffraction measurements to block the unscattered electron beam to obtain diffraction images with good contrast).

Next, to answer the second question, we reanalyzed our data by selecting regions of interest (ROIs) adjacent to the diffraction peaks (henceforth referred to as "shadow areas") and comparing them to the diffraction peaks themselves. We ran the analysis again, using the same methodology (explained in the SM of Hui et al. (*4*)) and plotted the results in Fig. 1. The background modulation signal of the shadow areas was found to have minimal amplitude (dashed blue line) compared to the diffraction oscillations at all diffraction orders (presented in solid black dotted connected with dashed black line, and the red line is the smoothing of the diffraction dynamic signal as a guide to eye). We notice that the background modulation extricated from the shadow areas next to the first-order peaks was slightly more than the second shadow areas because the first-order diffraction peaks are closer to the beam blocker, and this shows the opposite tendency compared with our signals.

Furthermore, in our dynamic diffraction signal, the third-order oscillation is more than the second-order oscillation, which in turn is more than the first-order diffraction. If Baum and Ropers' artifact hypothesis were correct, the diffraction signal would correspond to and scale proportionally with the background oscillations under all conditions. However, Baum and Ropers contradict their own assumption that we observed an interference artifact by attributing the varying oscillation levels at different diffraction orders to the "Debye-Waller effect." This explanation is inconsistent because the Debye-Waller effect, driven by thermal effects, typically occurs on a picosecond timescale—several orders of magnitude longer than the timescale of our measurements.

To evaluate the background oscillation impact, we subtracted the background modulation from the measured signal (in Fig. 1) and compared the results. Hence, the measured diffraction data, as presented in Fig. 5 of Hui et al. (*4*), were then plotted alongside the subtracted results (blue dotes connected with dashed blue lines) and shown in Fig. 2. The changes in the diffraction data are minimal, demonstrating the limited influence of background modulation on our measured diffraction signal.

*Our conclusion is that the background modulation does not noticeably affect the diffraction dynamics signal, as evidenced by the comparison of diffraction signals from the peaks and shadow regions (Fig.2).*

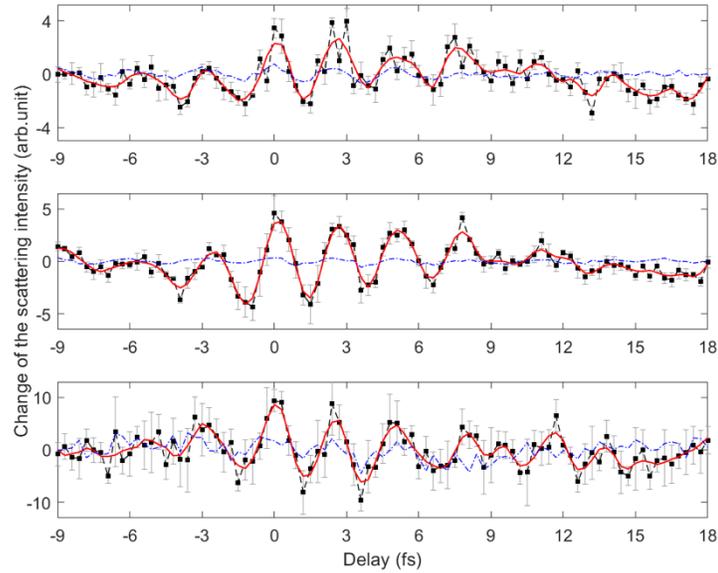

**Fig. 1. The measured diffraction dynamics signals for (from top to down) the 1st-, 2nd-, and 3rd-order peaks are shown as black dots connected by a dashed black line (as presented in Fig. 5 in Hui et al. Science Advances 2024) (*4*). A smoothed red line is included as a visual guide. These dynamics signals are compared to the background modulation signal (blue dashed line) obtained from the shadow regions adjacent to the 1st-, 2nd-, and 3rd-order diffraction peaks.**

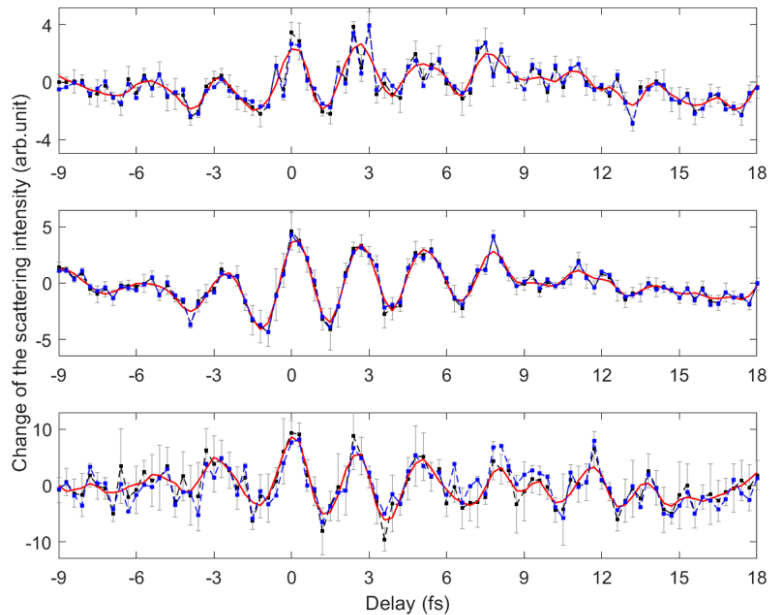

**Fig. 2. The measured diffraction dynamics signals for (from top to down) the 1st-, 2nd-, and 3rd-order peaks are shown as black dots connected by a dashed black line (as presented in Fig. 5 in Hui et al. Science. Advances 2024) (*4*). A smoothed red line is included as a visual guide. These dynamics signals are compared to the results obtained by subtracting the background modulation signal from the measured diffraction dynamics signals (shown in Fig. 1), represented by blue dots connected with a dashed blue line.**

Then we raised another question;

**(iii) what is the correlation between background modulation and diffraction dynamics?**

To answer this question and further ensure that our results were not influenced by the background modulation, we performed another measurement after reducing the pump pulse power —to a level that would not trigger significant intraband dynamics in graphene (1.75 V/nm in comparison to 2.4 V/nm used in our measured data in Fig. 5 in Hui et al. (*4*)). As shown in Fig. 3, we observed no diffraction dynamic signal (shown in black dotes connected with black dashed line) despite the presence of background oscillations in the shadow areas of diffraction peaks (presented in blue dashed line). This background oscillation in this measurement is at the same level of the background oscillations in our presented data in in Hui et al. (*4*) (we plot it for comparison in Fig. 3 in red dashed line). This observation rules out the possibility that the background modulation was responsible for the diffraction dynamics signal, or electron modulation due to interference. Furthermore, if the modulation had been due to any other interference effects, as Baum and Ropers suggest, it would have been evident in this scan.

**(iv) Why don't we see a diffraction dynamical signal at these low pump power measurements in Fig. 3?**

The primary reason for the diffraction dynamics signal observed in Fig. 5 in Hui et al. (*4*) is the field-induced intraband dynamics within the conduction band (CB), which becomes significant when the number of excited carriers is sufficiently high to generate a detectable signal. At lower field strengths, although the field still induces electron motion in the CB, the number of excited carriers is too small to produce an observable signal, as the carrier population increases nonlinearly with field strength.

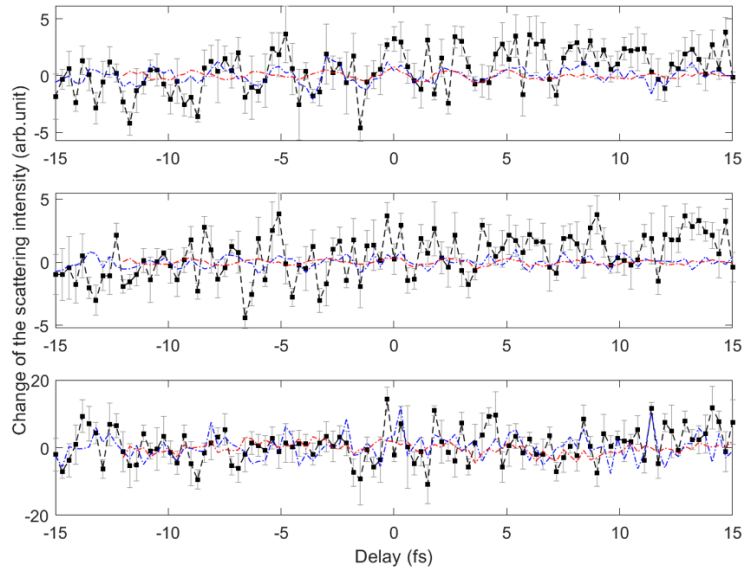

**Fig. 3. The measured diffraction signals for (from top to down) the 1st-, 2nd-, and 3rd-order diffraction peaks, pumped by a low-intensity laser pulse, are represented as black dots connected by a dashed black line. These signals are compared to the background modulation signals: (1) from the shadow regions adjacent to the 1st-, 2nd-, and 3rd-order diffraction peaks (blue dashed line), and (2) from Fig. 1 (red dashed line). No distinct diffraction dynamics are observed in the three orders, despite the background oscillation noise being comparable to the levels observed in the measured diffraction signals in Fig. 1.**

## 4. Addressing the Unrealistic Signal and Noise Levels

We generated electron pulses via photoemission using a 30 fs UV laser pulse. The estimated electron pulse duration based on classic space-charge calculations is ~300 fs. The gating time window, however, is 625 attoseconds. Consequently, only 0.2% of the electron beam (assuming a Gaussian distribution of the main electron pulse) is gated, with an estimated 5,000 gated electrons per image out of a total of 2.5 million electrons per image. This is sufficient for probing dynamics and our results averaged over seven scans. Worth mentioning, we added beam stabilization systems for both pump and probe laser beams, covered the entire optical path, to avoid any interruption duration the data acquisition, etc. to improve the signal to noise ratio.

Notably, our laser system has an optimized beam stability of 0.3% (root mean square) and other advanced noise reduction techniques. These include low-jitter gating and efficient background subtraction, as detailed in the supplementary materials. These measures effectively mitigate noise contributions, reinforcing the reliability of the reported results.

The critique's calculation of expected Bragg spot intensity changes relies on a simplistic model as in traditional atomic dynamics, which neglects the fact that electron momentum doesn't have a uniformly defined shape and evolves differently as a function of time

Regarding the signal amplitude, the laser pulse gating the electrons may amplify coherence and spatial focusing effects on the gated portion of the beam. These effects enhance the likelihood of producing a stronger or more distinct diffraction pattern, despite the gated electrons comprising only a small fraction of the total pulse. Coherence and preferential alignment could further intensify their interaction with the sample, yielding a more pronounced diffraction signal.

In our experiment, after scattering from the graphene sample, the ungated electrons in the main beam are largely blocked by a beam stopper, leaving only a small fraction for reference diffraction images. We normalize the gated electron signal relative to this reference fraction, which has a count comparable to the diffraction peaks, correcting for UV and electron count fluctuations. This approach minimizes experimental jitter and drift. The relative change in gated electron intensity, attributable to graphene dynamics, is calculated by normalizing to pre-arrival data (-9 to -8 fs) and presented as a percentage change. This normalization focuses on the gated electrons' contributions, explaining the use of arbitrary units (arb. un.) on the y-axis. *Finally, it is important to point out that in Hui et al. (4), we did not interpret the quantitative diffraction oscillation amplitude.*

## 5. Terminology Debate: "Attosecond Electron Microscopy"

Next, Baum and Ropers object to our use of the term "microscopy" in Hui et al. (2024) (4), where we referred to our experiment as attosecond electron microscopy ("attomicroscopy"). Their objection is based on their definition of microscopy, which they describe as *"the production of magnified images of a material in spatial coordinates, a result that is not achieved."*

However, this definition is inconsistent with established uses of the term. Scanning tunneling microscopy (STM), for example, does not produce magnified images in the traditional sense but still falls under the microscopy umbrella.

The etymology of *microscopy* (Greek: *mikros* = "small," *skopein* = "to examine") encompasses techniques beyond magnified imaging. In transmission electron microscopy (TEM), techniques such as real-space imaging, diffraction imaging, and electron energy loss spectroscopy (EELS) are all classified as microscopy. In Hui et al., 2024 (*4*), we achieved attosecond temporal resolution using TEM's diffraction imaging mode, with potential applications across other TEM techniques.

Thus, calling our method *attosecond electron microscopy* (or *attomicroscopy*) is both accurate and appropriate.

Moreover, the work demonstrated by Ropers's group in Gaida et al. (Nature Photonics 2024)(*2*), presents electron energy spectral mapping based solely *on electron spectral modulation* and the author titled the work **"Attosecond electron microscopy by free-electron homodyne detection"** despite his own definition mentioned in the comment about the definition of "microscopy" imaging.

**6. Lack of due diligence:**

Baum and Ropers seem to have overlooked critical details. For instance, we specified the optical focus diameter (~200 μm) on page 2 of Ref. 4, while the electron pulse beam size is ~80 μm. We also provided comprehensive data, including raw measurements and the integration times attached. Our electron pulse duration estimates between ~300 fs account for space charge effects and the photoemission geometry (41.5 cm from photocathode to sample).

We perform typical ultrafast electron diffraction measurements to find time-zero. Then, we determine the exact time zero from the dynamic curve. Next, we adjust the temporal overlap between the pump and the optical gating pulse to ensure the optical gating pulse is in perfect overlap with the main electron pulse. Then, we performed the attosecond electron diffraction time-resolved measurements.

Worth notes, after the draft of Ref. 4 was uploaded to Arxiv platform on May 4, 2023, Baum and Ropers contacted us the very next day, asserting that our results is caused by electron beam deflection akin to their earlier work (Ref. (*6*)), and not attosecond electron dynamics measurements. Although this presumption is absent from their commentary, Baum and Ropers now have access to our raw data and are encouraged to revisit these conclusions in light of the raw data provided (7).

**7. Conclusion: Refuting the Claims of Baum and Ropers**

Based on the experimental evidence presented above, we conclude that the background modulation observed in our measurements has a minimal effect on the diffraction dynamics signals and does not alter the conclusions of our study. **The data presented in Hui et al. Science Advances 2024)** (*4*). **provides clear evidence of graphene's intraband dynamics, unaffected by optical interference artifacts.** The concerns raised by Baum and Ropers regarding electron gating, modulation principles, signal and noise level, and optical interference are based on misunderstandings or misinterpretations of our experimental setup. Our findings, supported by extensive experimental confirmation, and data analysis, decisively demonstrate that the diffraction dynamics observed in our study are not artifacts but represent genuine ultrafast electron motion dynamics in graphene.